\documentclass{article}

\usepackage{graphicx}
\usepackage{cite}
\usepackage{amsmath}
\usepackage{url}
\usepackage{lscape}
\usepackage{indentfirst}
\usepackage{latexsym}
\usepackage{multirow}
\usepackage{tabls}
\usepackage{wrapfig}
\usepackage{longtable}
\usepackage{supertabular}
\usepackage{caption}
\usepackage{subcaption}
\usepackage{amsmath}
\usepackage{amssymb}
\usepackage{setspace}
\usepackage[bookmarks, colorlinks=true, plainpages=false, citecolor=blue, urlcolor=blue, filecolor=blue, linkcolor=blue]{hyperref} 
\usepackage[font={footnotesize,stretch=1.0}]{caption}

\newcommand{\pd}[2]{\frac{\partial #1}{\partial #2}}

\usepackage[margin=1.0in]{geometry}
\usepackage[utf8]{inputenc}
\usepackage[T1]{fontenc}

\title{Analytic model for neutral penetration and plasma fueling}

\author{George J. Wilkie}

\begin{document}

\maketitle

\begin{abstract}
   Neutral atoms recycled from wall interaction interact with confined plasma, thereby refueling it, most strongly in the region closest to the wall. This occurs near the X-point in diverted configurations, or else near the wall itself in limited configurations. A progression of analytic models are developed for neutral density in the vicinity of a planar or linear source in an ionizing domain. First-principles neutral transport simulations with DEGAS2 are used throughout to test the validity and limits of the model when using equivalent sources. The model is further generalized for strong plasma gradients or the inclusion of charge exchange. An important part of the problem of neutral fueling from recycling is thereby isolated and solved with a closed-form analytic model. A key finding is that charge exchange with the confined plasma can be significantly simplified with a reasonable sacrifice of accuracy by treating it as a loss. The several assumptions inherent to the model (and the simulations to which it is compared) can be adapted according to the particular behavior of neutrals in the divertor and the manner in which they cross the separatrix.
\end{abstract}

\section{Introduction}

Reliable prediction of neutral transport in magnetic confinement is critical for interpretation of line radiation diagnostics \cite{stotler_examination_2020,maan2024estimates,wilkie2024reconstruction}, divertor detachment \cite{dudson_role_2019,krasheninnikov_divertor_2016}, and fueling \cite{mordijck2020overview,haskey2024near}. Except in regions with especially dense gas, kinetic theory is necessary to capture the free-streaming of neutrals between charge exchange events and before ionizing. 

The most flexible and robust method for this prediction is the Monte Carlo method. EIRENE \cite{reiter2005eirene} and DEGAS2 \cite{stotler_neutral_1994} are the leading tools for doing so, and their fundamental algorithms were adapted from techniques for neutron transport prediction \cite{spanier_monte_2008}. Modern alternatives continue to be developed and improved, including Eunomia \cite{chandra2021b2} and AtOMC \cite{wilkie2025demonstration}. The latter is a recently-benchmarked adaptation of OpenMC for atomic physics, and is still in early stages of development. Deterministic Boltzmann solvers also continue to be improved \cite{holland2022updating,bernard2022kinetic,wilkie2023multidisciplinary}. KN1D \cite{labombard2001kn1d} is a notable example of a one-dimensional kinetic neutral transport solver that is useful for interpreting experimental diagnostics and continues to be modernized \cite{holland2022updating}.

Direct simulation remains the most reliable prediction of atomic and molecular transport and the resultant plasma fueling. Reduced models, however, are useful for rapid interpretation of experimental data, verification of computational models, and basic physics understanding. Neutral penetration has recently been characterized with exponential scaling relations \cite{fujii2025scaling}.

In this work, a reduced analytic model is developed from first-principles neutral kinetic theory. It is based on a relatively simple solution to the kinetic equation under a loss, source, and free-streaming. While the distribution function is readily written in closed-form, it does not admit closed-form moments except with very particular forms of the source velocity distribution. An asymptotic technique is employed to find the approximate spatial distribution of neutral density, $n_n$ in a strongly ionizing, or otherwise ``lossy'' medium. In the most basic analysis, it will be found that $n_n \propto r^{k}\exp\left[- \left(\gamma r / 2 v_{tn}\right)^{2/3} \right]$, where $r$ is the distance from the source, $\gamma$ is the loss rate in units of inverse time, and $v_{tn}$ is the thermal velocity of a Maxwellian source. The power of the additional prefactor is $k=-1/3$ for a planar source and $k=-1$ for a linear source. As additional complications (spatial dependence and charge exchange) are included, this fundamental form will be modified either by replacing $\gamma$ with a more generalized function or by adding additional terms with different prefactors and arguments of a similar exponential.

This model is most applicable in the free-molecular flow (high Knudsen number) limit. In regimes where atoms undergo many scattering-like events (including charge exchange) before ionizing, a random walk estimate for scattering is more appropriate. In this limit, a fluid approach is more rigorous and a diffusion operator rigorously captures the scattering process \cite{goldston2020introduction}. While the Monte Carlo algorithm is designed to handle the additional complication with frequent scattering events, albeit at the expense of increased computational cost, the analysis here gets progressively more complicated and only the ``first generation'' charge-exchanged atoms are considered. It will be found, at least for the domain and source considered here, even this may be unnecessary, and considering charge exchange as a loss inside the separatrix gives reasonably accurate results compared to simulation.

\subsection{Simulation Setup}

DEGAS2 is an atomic and molecular transport solver primarily for plasma-neutral interaction calculations \cite{stotler_neutral_1994}. It solves the Boltzmann equation for neutral atomic and molecular test particles under a large collection of reactions with plasma species, each other, and with plasma-facing components. Tallies (e.g. estimates for neutral density and ionization rate) are collected in finite volumes bounded by quadratic surfaces. Within these volumes, all background properties are constant. For these simulations, the background plasma consists of deuterium ions and electrons, with deuterium atoms as the test particle species.

\begin{figure}
   \begin{center}
      \includegraphics[width=0.5\textwidth]{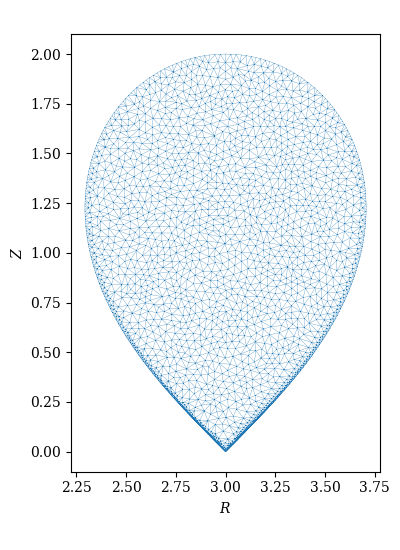}
      \caption{\label{lemngeom}An illustration of the unstructured mesh used for DEGAS2, as bounded by a lemniscate.}
   \end{center}
\end{figure}

For this work, the domain is isolated to the confined intra-separatrix plasma. In Section \ref{planeSec}, the domain is a box with planar symmetry and an extent along the $x$-direction of 60 cm, divided into 50 volumes within which tallies are estimated (and, in Section \ref{ped1dSec}, between which the plasma varies). In the 2D axisymmetric separatrix-like configuration, the boundary is a linearly-segmented lemniscate in the poloidal plane with an X-point at a major radius of 3 m, and a vertical minor radius of 1 m. This domain is divided into an unstructured triangular mesh with about 5000 elements with some intentional concentration near the X-point. See Figure \ref{lemngeom}. An additional very small volume is added adjacent to the X-point to simulate a linear source around the X-point. The boundary of the lemniscate is transparent: all neutral trajectories that reach it are lost.

The X-point source has a magnitude of $10^{22}$ particles per second, while the planar source is $10^{22}$ particles per $\mathrm{m}^2$ per second. For all these simulations, one million ``flights'' (trajectory samples) were used. The source velocity distribution is an isotropic Maxwellian at $T_n = 100 \mathrm{eV}$. It will be found that the appropriate normalized loss rate includes a factor of thermal speed, so by varying the loss rate, variation the source temperature is likewise covered.

Throughout these analyses and simulations, electron-impact ionization is an included reaction \cite{bray_electron-_2012}. It is built from a collisional radiative model treating the ground state as the only transporting species and excited states are averaged upon to find an effective ionization rate, which is tabulated as a function of electron density ($n_e$) and temperature ($T_e$). If $\sigma_{iz}$ is the cross section for ionization, the reaction rate is denoted $\left\langle \sigma_{iz} v \right\rangle$, with the angle brackets denoting an average over the electron velocity distribution. In this work, the ionization reaction will typically be characterized in terms of the loss rate including a factor of electron density:
\begin{equation} \label{lossdef}
   \gamma_{iz} \equiv n_e \left\langle \sigma_{iz} v \right\rangle.
\end{equation}
Similarly, the charge exchange reaction \cite{krstic_elastic_1998} is included for some cases, with $\gamma_{cx} = n_i \left\langle \sigma_{cx} v \right\rangle$ with ion density $n_i = n_e$. Generally, and in DEGAS2, this rate depends on the neutral velocity since the ion and atom velocities are coupled through the cross section $\sigma_{cx}$ (whereas electron velocities are well separated from the atom velocity). For the purposes of the model, the rate is queried from DEGAS2's table using $v_{tn}$ as the relative speed of neutrals with respect to the ions' rest frame.

The remainder of this paper is organized as follows. Section \ref{uniformplaneSec} treats the simplest and most pedagogical case of a planar neutral source in a uniform ionizing domain, wherein the fundamental technique is detailed. Then, Section \ref{ped1dSec} considers a strongly-varying ionizing domain and a similar, though more complicated model function results for neutral density. The domain is then reconfigured to that resembling a generic magnetic separatrix, again first treating the plasma as uniform and ionizing in Section \ref{uniform2dSec}. Finally, a model charge exchange operator is added and the results are compared to more complete simulations in Section \ref{CXSec}. A summarizing and forward-looking discussion concludes in Section \ref{discussionSec}.

\section{Planar source} \label{planeSec}

This section is intended to introduce the analysis that is built upon later. It is most directly appropriate for fueling where neither the neutral source crossing the separatrix has significant poloidal dependence such as for main-chamber recycling \cite{whyte2005magnitude}.

The domain is simplified to a box with 1D dependence along the $x$ axis. First, a uniform lossy domain is considered, then the local loss rate is given a strong $x$-dependence to mimic a plasma edge transport barrier or ``pedestal''.

\subsection{Uniform loss with planar symmetry} \label{uniformplaneSec}

The simplest case that nevertheless exhibits key features of the analysis to follow is ballistic transport from a source at the boundary through a 1D domain with a uniform loss rate. Let $f(x,v)$ be the phase space distribution of neutrals produced with a velocity distribution $S(v)$ and eliminated at a loss rate $\gamma$ (in units of $\mathrm{s}^{-1}$). The kinetic equation describing this case is:
\begin{equation} \label{kinetic1d}
   v \pd{f}{x} + \gamma f = S(v) \delta(x),
\end{equation}
with $\delta(x)$ representing the Dirac distribution. With an integrating factor, Eq. \eqref{kinetic1d} admits an analytic solution: 
\begin{equation} \label{soln1dUniform}
   f(x,v) = \frac{S(v)}{v} e^{-\gamma x/v}.
\end{equation}
While the distribution function can be written in closed-form, its density moment $n_n \equiv \int f \,\mathrm{d}v$ cannot. This is of chief concern since the local ionization rate density $R_{iz}$ (and other important quantities of interest) are directly proportional thereto: $R_{iz} = n_n \gamma_{iz}$. In order to perform such an integration, one must choose a velocity dependence of the source. Throughout this work, the choice is an isotropic Maxwellian:
\begin{equation} \label{maxwsource}
   S(v) =\frac{2 S_0}{\sqrt{\pi} v_{tn}} e^{-v^2/v_{tn}^2},
\end{equation}
where $v_{tn}$ is the thermal speed associated with the source, $S_0$ is the rate of particles produced per unit time per unit area, the other two velocity coordinates are already averaged upon, and $S(v)$ is only defined for $v>0$. In this case, the density moment is:
\begin{equation} \label{nn1dUniform}
   n_n =\frac{2 S_0}{\sqrt{\pi} v_{tn}} \int\limits_0^\infty v^{-1} \exp\left[-\frac{v^2}{v_{tn}^2} - \frac{\gamma x}{v} \right] \,\mathrm{d}v.
\end{equation}
To approximate integrals of this type, this work will use the saddle point method, taking advantage of the relatively large loss rate $\gamma \gg v_{tn}/x$. In Eq. \eqref{nn1dUniform}, the integrand  has a saddle point at $v_* = v_{tn} \left(\gamma x / 2 v_t \right)^{1/3}$, and the integral in the asymptotic limit is:
\begin{equation} \label{nn1dUniformApprox}
   n_n(x)  \approx \frac{2 S_0}{\sqrt{3} v_{tn}} \left( \frac{\gamma x}{2 v_{tn}} \right)^{-1/3} \exp\left[ - 3 \left( \frac{\gamma x}{2 v_{tn}} \right)^{2/3} \right].
\end{equation}

To test the applicability of Eq. \eqref{nn1dUniformApprox}, several cases are prepared with DEGAS2 using electron-impact ionization as the only reaction included. For the purposes of estimating neutral density, this reaction is purely a sink precisely as it acts in Eq. \eqref{kinetic1d}, and DEGAS2 directly solves along trajectories by sampling the source distribution and dividing the domain into many sub-volumes to estimate neutral tallies. In these cases, the neutral thermal velocity corresponds to a temperature $T_n = m_n v_{tn}^2 / 2 = $ 100 eV. 
The electron density and temperature were chosen so that the three cases correspond to, within three significant figures, $\hat{\gamma} = (3, 30, 100)$, where $\hat{\gamma} \equiv \gamma / v_{tn}$ is an inverse ``mean free path''. This is a known to be an important quantity for fueling \cite{fujii2025scaling} and will remain dimensional with units of inverse length to stress that it does not directly scale with device size \cite{mordijck2020overview}. Table \ref{uniformtable} lists the properties of these test cases.
\begin{table}
      \caption{\label{uniformtable}List of the parameters used for the uniform-plasma cases considered here. U1-U3 are the planar symmetry cases, and X1-X3 are the X-point cases. Also shown are the normalized rates for ionization  and charge exchange, respectively.}
   \begin{center}
   \begin{tabular}{l|llll|llllll}
      Label & $\hat{\gamma}_{iz} (\mathrm{m}^{-1})$ & $\hat{\gamma}_{cx}(\mathrm{m}^{-1})$ & $n_{e} (\mathrm{m}^{-3})$ & $T_{e}$ (eV) & $T_i$ (eV)\\
      \hline
      U1 & 3.0 & \textemdash & $3.73\times10^{18}$ & 200 & \textemdash \\
      U2 & 30.0 & \textemdash &  $8.62\times10^{19}$ & 500 & \textemdash \\
      U3 & 100.0 & \textemdash &  $3.01\times10^{20}$ & 1000 & \textemdash \\
      \hline
      X1 & 0.986 & 1.52 & $3\times10^{18}$ & 200 & 300 \\
      X2 & 3.29 & 5.89 & $10^{19}$ & 300 & 500 \\
      X3 & 35.3 & 72.1 & $10^{20}$ & 500 & 1000 \\
   \end{tabular}
   \end{center}
\end{table}

\begin{figure}
   \begin{center}
      \includegraphics[width=1.0\textwidth]{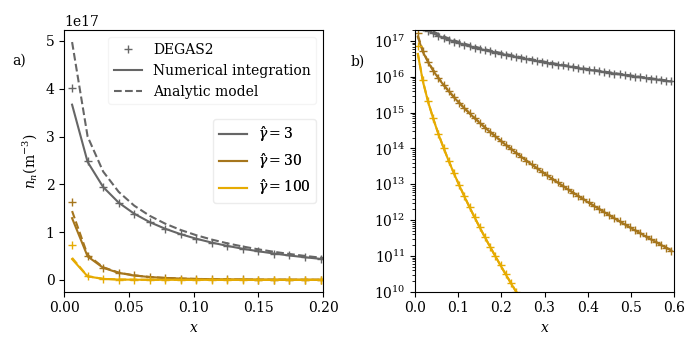}
   \end{center}
   \caption{\label{uniformn1d} Predicted neutral density through a uniform plasma domain with several normalized ionization loss rates $ \hat{\gamma} = n_e \left\langle \sigma_{iz} v \right\rangle/ v_{tn}$. On the left (a) is a linear scale at low $x$,  and on the right (b) is a logarithmic scale. The solid line are the results from DEGAS2, the crosses represent the numerical evaluation of the integral \eqref{soln1dUniform}, and the dashed line is the analytic closed-form approximation \eqref{nn1dUniformApprox}. Cases correspond to those labelled U1, U2, and U3 in Table \ref{uniformtable}.}
\end{figure}

Figure \ref{uniformn1d} shows these three cases comparing the DEGAS2 results to the model of Eq. \eqref{nn1dUniformApprox}. 
Also shown are the results from direct numerical calculation of the integral in Eq. \eqref{nn1dUniform}. 
Agreement between the model and simulation is generally good, with some exceptions near $x=0$. Since this is the foundation upon which more general models are built, it is worth examining the discrepancy. DEGAS2 utilizes ``suppressed ionization'' whereby it directly uses the analytic solution of $f$ in Eq. \eqref{soln1dUniform}, so it should, in principle, exactly match the direct numeric integration of Eq. \eqref{nn1dUniform}. The reason any disagreement between DEGAS2 and numerical integration is seen at at all in Fig. \ref{uniformn1d} is because DEGAS2 calculates a volume-averaged tally, whereas the function \eqref{soln1dUniform} is integrated at the particular values of $x$. Since $n_n$ has a strong second derivative near $x=0$, the volume-averaged density in a cell is slightly more than the value evaluated at the midpoint of the cell. A similar issue was encountered in the benchmarks of \cite{bernard2022kinetic} requiring DEGAS2 to use higher resolution to capture higher-order variability within Gkeyl spatial cells. Additional disagreement is seen between the closed-form saddle point solution \eqref{nn1dUniformApprox} and both the DEGAS2 results and direct integration of the kinetic equation. This disagreement is due to the breakdown of the saddle point approximation itself, wherein $\hat{\gamma}$ is not reliably considered asymptotically large, especially near $x=0$. Even so, a disagreement of only $\sim 25\%$ when the ``large'' asymptotic parameter is much smaller that unity (about 0.03 for the $\hat{\gamma} = 3$ case at $x \approx 0.01$) is remarkably fortunate. This might be an indication that the analytic approximation of Eq. \eqref{nn1dUniform}, and those like it that follow, can be derived by other means.

\subsection{Pedestal loss with planar symmetry} \label{ped1dSec}

The solution to Eq. \eqref{kinetic1d} is readily found for any $\gamma(x)$ that admits a closed-form indefinite integral, but this work shall restrict itself to a particular functional form:
\begin{equation} \label{tanhprofile}
   \gamma(x) = \gamma_0 + \frac{\gamma_\infty - \gamma_0}{2} \left[ 1 + \tanh\left(\frac{x-x_0}{\Delta} \right) \right], 
\end{equation}
where $x_0$ and $\Delta$ are parameters defining the shape of $\gamma(x)$, while $\gamma_0$ and $\gamma_\infty$ are the left and right asymptotes of $\gamma$, corresponding to the SOL and core asymptotes, respectively. This form is inspired by the plasma profiles typically fit in the pedestal region of tokamaks \cite{groebner1998scaling}. The interpretation of $\Delta$ and $x_0$ are the ``pedestal''
width and the location of largest gradient, respectively. 
The loss profile shapes that will be used are listed in Table \ref{pedtable}. Note that the shapes and asymptotes of the internally-fit loss-rate profile do not necessarily match that of the plasma itself, since the former is a tabulated function of the latter.

Under the functional form of Eq. \eqref{tanhprofile}, the solution to \eqref{kinetic1d} is:
\begin{equation} \label{fsoln}
   f(x,v) = \frac{S(v)}{v} \exp\left[ \frac{x_0 - x}{v}\frac{L_0 + L_\infty}{2} \right] \left[ \frac{ \cosh\left( \frac{x - x_0}{\Delta} \right)}{\cosh\left( \frac{x_0}{ \Delta} \right)} \right]^{\frac{\Delta}{v}\left( L_\infty - L_0 \right)}. 
\end{equation}
The density moment under the same approximation as the previous section is:
\begin{equation} \label{pedn1d} 
   n_n(x) \approx \frac{2 S_0}{\sqrt{3 \pi}} \left( \frac{u(x)}{2} \right)^{-1/3} \exp\left[- 3 \left( \frac{u(x)}{2} \right)^{2/3}\right],
\end{equation}
where:
\begin{equation}
   u(x) =  \Delta \frac{L_\infty - L_0}{2}\ln \left[ \frac{\cosh\left( \frac{x -x_0}{\Delta} \right)}{ \cosh\left( \frac{x_0}{\Delta} \right)}\right] + \frac{L_0+L_\infty}{2}\left(x-x_0\right).
\end{equation}

\begin{figure}
   \begin{center}
      \includegraphics[width=1.0\textwidth]{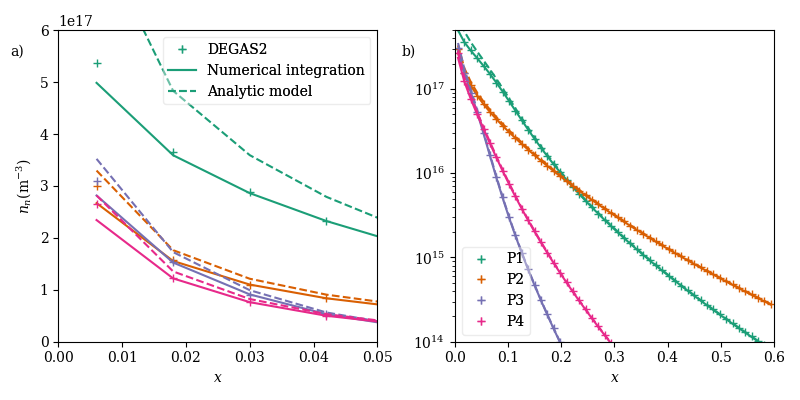}
   \end{center}
   \caption{\label{ped1d} Comparison between DEGAS2 (solid line) using constructed pedestal-like profiles versus the analytic model of Eq. \eqref{pedn1d} (dashed line) and direct numerical integration of Eq. \eqref{fsoln} (crosses). The four colors correspond to the four pedestal shapes listed in table \ref{pedtable}.}
\end{figure}

With this model, the neutral density throughout a medium with a spatially-varying loss rate can be estimated. Again, DEGAS2 is utilized for testing. Four artificial pedestals are constructed with the properties listed in Table \ref{pedtable}. 
\begin{table} 
   \begin{center}
   \begin{tabular}{l|llll|llllll}
      Label & $\hat{\gamma}_0$ & $\hat{\gamma}_\infty$ & $ \Delta$ & $x_0$ & $n_{e0}$ & $n_{e\infty}$ & $T_{e0}$ (eV) & $T_{e\infty}$ (eV) & $\Delta_e$ & $x_{0e}$\\
      \hline
      P1 & 0.429 & 13.7 & 0.0890 &  0.0537 & $3\times10^{18}$ & $5\times10^{19}$ & 10 & 1000 & 0.05 & 0.1 \\
      P2 & 7.43 & 9.29 & 0.130 & 0.0827 & $2\times10^{19}$ & $5\times10^{19}$ & 100 & 3000 & 0.1 & 0.1 \\
      P3 & 4.0 & 42.4 & 0.0444 & 0.0313 & $10^{19}$ & $2\times10^{20}$ & 100 & 3000 & 0.03 & 0.05  \\
      P4 & 7.12 & 25.8 & 0.0439 & 0.0504 & $2 \times10^{19}$ & $2 \times 10^{20}$ & 300 & 10000 & 0.05 & 0.05\\
      \hline
   \end{tabular}
   \end{center}
   \caption{\label{pedtable}List of the parameters determining the loss rate profiles tested for the spatially-varying planar cases. $\Delta_e$ and $x_{0e}$ correspond to the pedestal shape properties of the electron population, where $\Delta$ and $x_0$ are the parameters for the fitted shapes of the ionization loss rate profile $\gamma(x)$.}
\end{table}
In DEGAS2, electron density and temperature profiles are constructed similarly to Eq. \eqref{tanhprofile}. Then, the ionization rate throughout the domain is fit to Eq. \eqref{tanhprofile} to find the listed properties that determine $\hat{\gamma}(x)$. The results from these simulations are shown in Fig. \ref{ped1d}. 
Similar agreement is seen as for the uniform case: direct integration agrees better with the DEGAS2 than the model, especially at relatively low $\hat{\gamma}$. Significant disagreement is seen for the ``P1'' case at small $x$ since the ``large'' asymptotic parameter is as low as $5\times 10^{-3}$. Agreement is recovered deeper into the domain as $\hat{\gamma} x$ becomes larger.

\section{Radial X-point source}

In diverted magnetic geometry, recycling neutrals are present at highest density near the divertor. Since the magnetic null (``X-point'') is typically the closest part of the confined region to the divertor, that is also where the confined plasma can be expected to interact most strongly with recycling neutrals. Reflection from the walls and interaction with the scrape-off layer plasma significantly complicates the picture necessitating first-principles kinetic simulation for accurate fueling profiles. In this section, the extreme case is considered of neutrals crossing the separatrix \emph{only} at the X-point. The solution associated with this $\delta$ function source can in principle be extended as a Greens' function for a more general poloidal dependence of neutrals entering the domain from outside the separatrix.

In this section, the geometry of the problem is adjusted to one of radial symmetry and a linear source of neutrals localized at the X-point (which subtends a circle of radius $R$). 
See Figure \ref{lemngeom} for an illustration of the unstructured mesh for these cases. 

The source of neutrals is again an isotropic Maxwellian at 100 eV, nearly point-wise in the poloidal plane (actually, a very small volume adjacent to the X-point and subtended around the ``tokamak'' axis of symmetry).

\subsection{Uniform lossy plasma} \label{uniform2dSec}

\begin{figure}
   \begin{center}
      \includegraphics[width=1.0\textwidth]{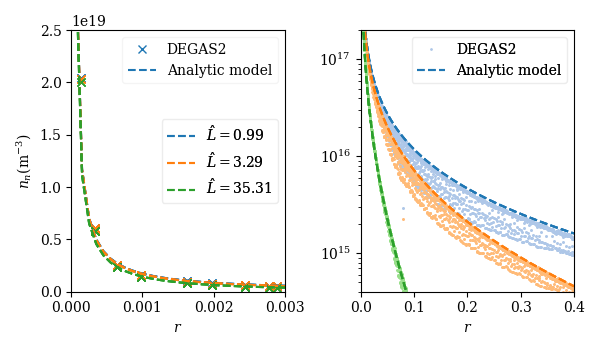}
   \end{center}
   \caption{\label{uniformn2d}Comparing DEGAS2 simulation results to the closed-form model of Eq. \eqref{nasymp} for three different uniform loss rates with a linear source. Neutral density is plotted versus distance to the X-point for each volume element in the mesh. The three cases are listed as X1, X2, and X3 in Table \ref{uniformtable}.}
\end{figure}

In coordinates where $r$ is the distance to the X-point and $v$ is the corresponding radial velocity, the kinetic equation is:
\begin{equation} \label{kineqx}
   v \pd{f}{r} + \gamma f = \frac{S(v)}{2 \pi R} \frac{\delta(r)}{r},
\end{equation}
which admits the solution:
\begin{equation} \label{xsoln}
   f(r,v) = \frac{S(v)}{2 \pi R r v_{tn} } \frac{1}{v} e^{- \gamma r/v}.
\end{equation}
With the now-familiar procedure of treating the source as Maxwellian and applying the method of steepest descent, the analytic model for neutral density is:
\begin{equation} \label{nasymp}
   n(r) \approx \frac{S_0}{\sqrt{12 \pi^3} R v_{tn} r} \exp\left[ - 3 \left( \frac{\gamma r}{2 v_{tn}} \right)^{2/3} \right].
\end{equation}
In this geometry, the neutral density takes a slightly different form owing to the cancellation of the factor of $v$ due to the velocity integration measure $\int \left[\ldots \right] v\mathrm{d}v$. This change is not inconsequential, as the solution more clearly departs from purely exponential behavior \cite{fujii2025scaling}. See Figure \ref{uniformn2d} for sample cases with three different pairs of uniform electron density and temperature (as listed in Table \ref{uniformtable}).

Note that the DEGAS2 results exhibit significant spread, with Eq. \eqref{nasymp} well-capturing the upper bound. The reason for this spread is a combination of: poloidal curvature of the lemniscate, toroidal curvature, and finite extent of the numerical source. These effects are beyond the scope of the analytic model, but are worth noting and some aspects are discussed further in Section \ref{discussionSec}.

Considering the spatial dependence of the neutral density from simulation, one can be easily imagine attributing the departure from exponential behavior to charge exchange, but this reaction is not included in these cases. This is addressed in the following section.

\subsection{Inclusion of charge exchange} \label{CXSec}

In general, charge exchange involves the inclusion of a collision integral in the kinetic equation. However, it is often useful to approximate it as a BGK operator \cite{wersal_first-principles_2015,bernard2022kinetic} of the form $\gamma_{cx} \left[ \left(n/n_i\right) f_i - f \right)$, where $f_i$ is a Maxwellian distribution of ions with density $n_i$ and thermal velocity $v_{ti}$. Effectively, this operator acts a sink of neutrals and replaces them with those sampled from the ion distribution $f_i$. It is exact only if ion velocities are decoupled from those of neutrals: either through a particular (unphysical) form of the cross section or by ion velocities being asymptotically large compared to neutrals \cite{wilkie2024reconstruction}. Therefore, this operator does not act on velocity space rigorously. One can imagine the approximation $v_{tn} \ll v_{ti}$ being valid, but it will not be formally applied here. The reason for not doing so is that neutrals penetrating the separatrix are likely to have already undergone a charge exchange reaction with a similar-temperature ion population outside the separatrix. If the neutral energy is very low compared to ions, they are not likely to have reached the separatrix in the first place.

With the BGK form of the charge exchange operator, the kinetic equation (with a Maxwellian source) reads: 
\begin{equation} \label{kineqcx}
   v \pd{f}{r} + \left(\gamma_{iz} + \gamma_{cx} \right) f =  \frac{S_0}{2 \pi^{2} R v_{tn}^2} e^{-v^2/v_{tn}^2} \frac{\delta(r)}{r} + \gamma_{cx} n(r) \frac{e^{-v^2/v_{ti}^2}}{\pi v_{ti}^2}.
\end{equation}
The ion population is taken to be spatially uniform. Eq. \eqref{kineqcx} introduces an additional complication: a source which depends on the undetermined neutral density. This remains tractable though by solving with a similar integrating factor:
\begin{align}
   f(r,v) =& \frac{ S_0 }{2 \pi^2 R v_{tn}^2 } \frac{1}{rv}\exp\left[ - \frac{v^2}{v_{tn}^2} - \frac{(\gamma_{iz} + \gamma_{cx})r}{v} \right] \nonumber \\
   &+ \frac{\gamma_{cx}}{\pi v_{ti}^2 v} \int\limits_0^r n(r') \exp\left[ -\frac{v^2}{v_{ti}^2} - \frac{\left( \gamma_{iz} + \gamma_{cx} \right) }{v} \left(r - r' \right)\right] \, \mathrm{d}r' 
\end{align}
Upon integrating over velocity (with the asymptotic solution of Eq. \eqref{nasymp}), the result is:
\begin{equation} \label{nwcx1}
   n(r) = n_0(r) + \frac{\gamma_{cx}}{\pi v_{ti}^2} \int\limits_0^r n(r')  \int  \exp\left[ -\frac{v^2}{v_{ti}^2} - \frac{\left( \gamma_{iz} + \gamma_{cx} \right) }{v} \left(r - r' \right)\right] \, \mathrm{d}v\mathrm{d}r', 
\end{equation}
where:
\begin{equation} \label{n0def}
   n_0(r) = \frac{S_0}{\sqrt{12 \pi^3} R v_{tn} r} \exp\left[ - 3 \left( \frac{ \left(\gamma_{iz} + \gamma_{cx} \right) r}{ 2 v_{tn}} \right)^{2/3} \right]
\end{equation}
is a slightly modified version of Eq. \eqref{nasymp} with $\gamma = \gamma_{iz} + \gamma_{cx}$, treating the charge exchange rate as an additional loss. The form of Eq. \eqref{nwcx1} is a Fredholm integral equation of the second kind. Formally, this is solved with a Neumann series\cite{spanier_monte_2008}, and this is the formal foundation upon which the Monte Carlo method is based. This is effectively a series of successive application of the integral kernel to an iterated $n(r')$, adding additional dimensionality to the integral at each iteration. A single such iteration is applied in this work, corresponding to neutrals undergoing only one charge exchange (``first generation''). After applying the asymptotic expansion on the velocity integral and substituting $n_0(r')$ in the integrand:
\begin{align} \label{nwcx}
   n(r) \approx n_0(r) +& \frac{\gamma_{cx} S_0}{\sqrt{12 \pi^3} R v_{tn}^2 } \frac{T_n}{T_i} \left( 1+ \frac{T_n}{T_i} \right)^{-1/6} \\
   &\times \left( \frac{ \left( \gamma_{iz} + \gamma_{cx} \right) r}{2 v_{tn}} \right)^{-1/3} \exp\left[ - 3 \left( \frac{ \left(\gamma_{iz} + \gamma_{cx} \right) r}{ 2 v_{tn}} \right)^{2/3} \left(1 + \frac{T_n}{T_i} \right)\right]. \nonumber
\end{align}

\begin{figure}
   \begin{center}
      \includegraphics[width=1.0\textwidth]{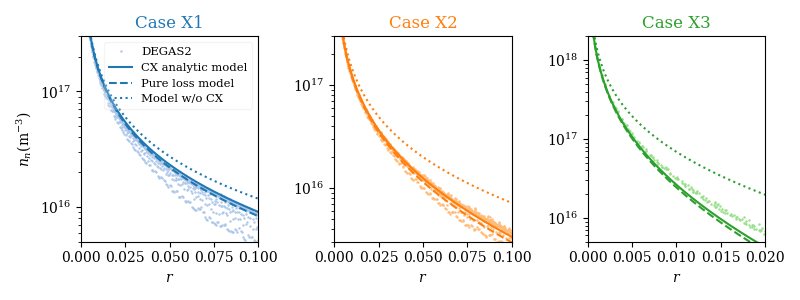}
   \end{center}
   \caption{\label{cxmodel2d}DEGAS2 results for neutral density in the mesh compared plotted versus distance of the centroid each element to the X-point (pale dots)with comparison to the analytic model of Eq. \eqref{nwcx} for the three different uniform-plasma cases. Also shown (as dotted lines) are the cases without charge exchange and (dashed lines) the prediction of Eq. \eqref{n0def} with charge exchange treated purely as a loss. Plasma parameters correspond to those listed as X1, X2, and X3 in Table \ref{uniformtable}.}
\end{figure}

Once again, the model Eq. \eqref{nwcx} is compared to results from DEGAS2 in Figure \ref{cxmodel2d} along with several other iterations for interpretation. These three cases are the same as in Fig. \ref{uniformn2d} (cases X1-3 in Table \ref{uniformtable}), but now with the charge exchange reaction included with ion populations at temperatures 300, 500, and 1000 eV, respectively. 
Several features of these cases are worthy of comment. Firstly, both the simulation results and model depart significantly from the case without charge exchange, so the effect is overall, as expected, nontrivial. Additionally, until the neutral density drops by nearly two orders of magnitude, both the model and the simulation are nearly indistinguishable from $n_0$ (Eq. \eqref{n0def}) which is labelled as the ``Pure loss model''. This indicates that the effect of charge exchange near the X-point is primarily as a loss mechanism. Finally, for the most extreme case X3, there remains an order-unity departure of the simulation results from the model(s) at $r \gtrsim 1 \mathrm{cm}$. There are three possible reasons for this discrepancy in terms of physics included in DEGAS2, but not the analytic model: full collision operator and cross section section for charge exchange rather than the BGK form; neutrals are permitted to undergo more than one charge exchange interaction in the simulation; and speed of neutrals with respect to the at-rest ion population may not be well captured by $v_{tn}$.

\begin{figure}
   \begin{center}
      \includegraphics[width=1.0\textwidth]{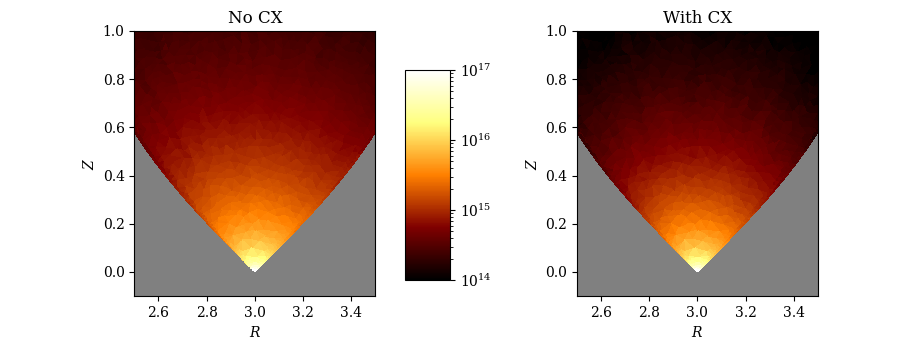}
   \end{center}
   \caption{\label{includecxcomp}Examples (corresponding to Case X1) of neutral density from DEGAS2 when ion charge exchange is ignored (left) or included (right).}
\end{figure}

The fact that charge exchange behaves as a loss near the X-point is evident from direct comparison of the simulation results in Figure \ref{includecxcomp}. 
There, it is evident that the neutral penetration is \emph{decreased} when charge exchange is included. This is due to the geometry of the problem. After a charge exchange event, the resulting neutral is sampled from the ion Maxwellian distribution. Most \textemdash about 75\% \textemdash of the post-charge exchange neutrals will therefore be on a trajectory that leaves the domain rather than penetrating deeper toward the core.

The lossy behavior of charge exchange observed here is not universal. In fact, the energy of the source neutrals is driven largely by the charge exchange reaction \emph{outside} the separatrix, manifesting as a larger $v_{tn}$ and a lower effective loss rate $\hat{\gamma} = \gamma / v_{tn}$.

\section{Discussion and conclusion} \label{discussionSec}

The progression of models presented in previous sections provides an analytic framework whereby penetration of neutrals within the separatrix can be transparently understood. Complications in various permutations were considered: planar versus linear source geometry, spatial variation of the plasma, and the inclusion of the BGK form of the charge exchange operator. 

The mean-free path, $\hat{\gamma}^{-1} = v_{tn}/\gamma$ remains an important parameter. However, since higher energy neutrals penetrate deeper than those of low energy, the neutral density does not decrease with a simple exponential even in a uniformly-ionizing domain. Instead, forms such as Eqs. \eqref{uniformn1d} and \eqref{uniformn2d} are more appropriate for general use and interpretation of spatially-resolved neutral diagnostics (once multiplied by the local emission rate). 

For the linear X-point source models, it was found that charge exchange acts primarily as a loss. Second-generation effects do appear inside the separatrix, but only sufficiently far from the X-point after the neutral density drops orders of magnitude. Additional results consistently reveal that treating charge exchange as a pure loss based primarily on reaction rates at the X-point is rather accurate when compared to more complete equivalent-source simulations, even in the presence of a plasma transport barrier. The reason is because of flux expansion: near the X-point, the gradient of a flux function ($\psi$) is small. So even if, for example, $n_e'(\psi)$ is very high, this does not necessarily translate to a very large $|\nabla n_e|$. Especially in cases where the neutral penetration depth is small, the plasma relevant for neutral interaction is effectively uniform in the vicinity of the X-point. This, coupled with the observation that the role of charge exchange is primarily as a loss mechanism, means that the relatively simple model of Eq. \eqref{uniformn2d} performs better than expected, even though it does not include a charge exchange reaction as such, nor does it allow for spatially-varying plasma. These additional complications can readily be included in a closed-form, albeit more complicated, model.

The model and analysis presented here opens up additional routes of inquiry. Primarily, how ought the neutral source near the X-point characterized in terms of their velocity and spatial distribution? How can the isotropic Maxwellian source be improved? Answering these questions will require follow-up study of the neutral transport outside of the confined plasma: in the scrape-off layer, divertor and private flux regions. 

One can imagine using the model in the form of Eqs. \eqref{uniformn2d} or \eqref{pedn1d} as a fitting function for neutral profiles observed experimentally. While developing the model further along these lines is encouraged, the reader is nevertheless urged caution until and unless the spatial and velocity distribution of neutrals is properly accounted for. Throughout the preparation of this work, it was easy to fit nearly any neutral profile observed in simulation to a function like \eqref{pedn1d}. Once such a function is fit to data, the physical relevance of the loss rate, the spatial or velocity distribution of the neutral source, or the role of the spatial coordinate $r$ can all be misinterpreted if one is not careful.

Throughout this analysis, the open magnetic field region has been excluded entirely. Instead, the neutral source was modelled as an isotropic Maxwellian at a given temperature. While $T_n = $ 100 eV was used universally throughout this work, sufficient generality was nevertheless included (with respect to testing the model) via a range of values of $\hat{\gamma}$ \textemdash loss rate normalized to an inverse-length via the neutral thermal speed. Similarly, device size (particularly major radius $R$) appears in a prefactor on the source rate, making the neutral density profile proportional to the number of source particles per length around the circumference of the magnetic null. 

The X-point fueling source presented here is just that: point-wise in the poloidal plane. Considering the very strong drop-off of the neutral density (e.g. within $r < 1 \mathrm{mm}$ in Fig. \ref{uniformn2d}a), a more general spatial distribution of neutrals is expected to significantly impact the results. Fortunately, the model source is a Dirac delta, so that the model itself can act as a Greens' function for follow-up work once the neutral spatial distribution incoming from the divertor is more properly characterized. Beyond the spatial distribution of neutrals, the isotropic Maxwellian source developed upon here is likely also insufficient.
Modifications such as a cosine velocity angle distribution or Poisson energy distribution can be readily incorporated, but further study outside the scope of this work is required. These results do, however, isolate the key characteristics needed for such a study: the neutral velocity distribution near the X-point, and the spatial distribution of the incoming flux.

The author is grateful to S. Haskey, Q. Pratt, A. Bortolon, and F. Parra for fruitful discussions.
This work was supported by the U.S. Department of Energy through under contract number DE-AC02-09CH11466 and the SciDAC project ‘Computational Evaluation and Design of Actuators for Core-Edge Integration’ (CEDA). The United States Government retains a non-exclusive, paid-up, irrevocable, world-wide license to publish or reproduce the published form of this manuscript, or allow others to do so, for United States Government purposes.

\bibliographystyle{unsrt}
\bibliography{zotero}

\end{document}